\documentclass[prd,aps,10pt,showkeys,nofootinbib,showpacs,twocolumn]{revtex4-2}

\usepackage{amsmath,amssymb,amsfonts,amsthm}
\usepackage{amsbsy} 
\usepackage{epsfig}
\usepackage{latexsym}
\input amssym.def
\input amssym.tex

\usepackage{hyperref}

\newcommand{\oarX}[1]{\href{http://arxiv.org/abs/#1}{{\ttfamily #1}}}
\newcommand{\arX}[1]{\href{http://arxiv.org/abs/#1}{{\ttfamily arXiv:#1}}}
\newcommand{\doin}[2]{\href{http://dx.doi.org/#1}{#2}}

\def\barr{\begin{array}}
\def\earr{\end{array}}

\def\ben{\begin{equation}}
\def\een{\end{equation}}
\def\bs{\begin{subequations}}
\def\es{\end{subequations}}
\def\bena{\begin{eqnarray}}
\def\eena{\end{eqnarray}}

\def\im{{\rm i}}

\def\be{\begin{equation}}
\def\ee{\end{equation}}

\def\bes{\begin{eqnarray}}
\def\ees{\end{eqnarray}}

\newcommand{\dd}{\mathrm{d}}

\begin{document}

\title{Frozen formalism and canonical quantization in group field theory}

\author{Steffen Gielen}
\affiliation{School of Mathematics and Statistics, University of Sheffield, Hicks Building, Hounsfield Road, Sheffield S3 7RH, United Kingdom}
\email{s.c.gielen@sheffield.ac.uk}
\date{\today}

%%%%%%%%%%%%%%%%%%%%%%%%%%%%%%%%%%%%%%%%%%%%%%%%%%%%%%%

\begin{abstract}
Canonical quantization of gravitational systems is obstructed by the problem of time. Due to diffeomorphism symmetry the Hamiltonian vanishes: dynamics with respect to a background time parameter appears ``frozen.'' Two strategies towards the quantization of such systems are the identification of a clock degree of freedom before quantization (deparametrization), and quantization on a kinematical Hilbert space which is subject to constraints (Dirac quantization). The usual canonical quantization in quantum field theory is analogous to deparametrization. Here we introduce a frozen formalism and Dirac quantization for a complex Klein--Gordon scalar field, and show that the resulting theory is equivalent to usual canonical quantization. We then apply the formalism to the group field theory formalism for quantum gravity, for which both deparametrization and a  ``timeless'' quantization have been proposed in past work. We show how a frozen formalism for group field theory links between these two existing approaches, and illustrate in particular the construction of physical observables. We derive effective cosmological dynamics for group field theory in the new formalism and compare these to previous work. The frozen formalism could be extended to other approaches to quantum gravity that do not use a preferred time parameter.
\end{abstract}

\keywords{canonical quantization; group field theory; problem of time; Dirac quantization}

\maketitle

\section{Introduction}

Canonical quantization, as presented in undergraduate textbooks on quantum mechanics, provides an in principle direct route from any classical to the corresponding quantum theory: starting from a classical theory defined by an action, one runs the Legendre transform to obtain the corresponding Hamiltonian theory, uses its Poisson structure to define canonical commutation relations and constructs quantum observables as Hermitian operators corresponding to classical phase space functions. In practice, ambiguities and additional choices appear for all but the very simplest systems motivating, {\em e.g.}, the more systematic approach of geometric quantization \cite{geometric}.

Things become more complicated for systems with gauge symmetries: not any phase space variable is now an observable, and one would like to focus on the dynamics of observables only. Moreover, gauge symmetries are associated to constraints which must be implemented in the quantum theory. If the gauge symmetry involves diffeomorphisms of time, gauge transformations and dynamics are intertwined; the attempt to define dynamics with respect to a given time parameter leads to a ``frozen formalism'' with vanishing Hamiltonian. A Hamiltonian formalism able to deal with gauge symmetries and in particular with diffeomorphism-invariant theories was developed by Dirac \cite{diracqu}. Constructing a quantum theory via Dirac's algorithm then leads to the infamous {\em problem of time} \cite{frozenrefs}: all states and observables are independent of the time parameter used to set up the theory. Evolution must then be defined in {\em relational} terms, as the evolution of some degrees of freedom with respect to others \cite{reldirc}.

In this paper we focus on two of the most popular approaches in the canonical quantization of generally covariant systems: {\em deparametrization} (or reduced quantization) in which one of the dynamical variables is identified as a ``clock'' before quantization, and {\em Dirac quantization} in which one first constructs a kinematical Hilbert space and demands that physical states satisfy the quantum version of the constraints of the theory\footnote{One can also define a notion of {\em quantum deparametrization} as, {\em e.g.}, in Ref.~\cite{trinitystuff}, where the clock is identified within a theory defined through Dirac quantization; we will not discuss this here.}. Deparametrization, while often easiest to implement, suffers from ambiguities and the lack of covariance, since it is not guaranteed that different choices of clock lead to equivalent theories \cite{deparamold}. Moreover, the degree of freedom used as a clock is often added by hand in order to guarantee its clocklike behavior; in quantum cosmology this is often a massless scalar field, which is classically monotonic on almost any solution. In contrast, Dirac quantization requires at least some control over the space of physical states (solutions to the constraint), and leads to a number of subtle technical issues \cite{ashtekartate}. The physical Hilbert space is usually not a subspace of the initial kinematical Hilbert space; it can be constructed through group averaging \cite{groupav}. The relation between the viewpoints of different clocks is clearer in Dirac quantization \cite{trinitystuff,switch}.

Most of the literature on deparametrization and Dirac quantization, and especially on comparisons between them, focuses on finite-dimensional systems such as particle models or homogeneous models in quantum cosmology. The model example on which the different approaches and challenges are often discussed is the relativistic particle in (Minkowski) spacetime, which has a one-dimensional diffeomorphism symmetry corresponding to reparametrizations on the worldline. Quantization of the relativistic particle leads to relativistic (Klein--Gordon) quantum field theory, where the wavefunction is promoted to a quantum field and the Hilbert space is enlarged to a Fock space of many-particle states. The usual presentation of canonical quantization of the Klein--Gordon field, however, follows the logic of deparametrization where a clock (here a time coordinate on Minkowski spacetime) is chosen before quantization. One has to ensure that the resulting formalism remains Lorentz covariant. There is however no remnant of the reparametrization invariance of the relativistic particle action.

Here we propose a frozen formalism for a complex Klein--Gordon quantum field: the quantum field is given an additional dependence on a proper time parameter $\tau$ but the equations of motion imply that this dependence is trivial. This allows defining a Dirac-type quantization in which there is a (Hamiltonian) constraint acting on a kinematical Hilbert space. We show how a physical inner product can be constructed through group averaging by extending the inner product for a relativistic particle to a Fock space. We construct both the physical Fock space and physical observables on this physical Fock space through ``projection'' maps that map operators on the kinematical Hilbert space to operators on the physical Hilbert space. For the Klein--Gordon field, we find that the resulting theory is then equivalent to the usual Fock quantization.

We then extend this formalism to the group field theory (GFT) approach to quantum gravity \cite{GFT}, which provides the main motivation for this work. GFT can be seen as a quantum field theory reformulation of the background-independent dynamics of quantum gravity as defined by spin foams \cite{SFGFT} and loop quantum gravity (LQG) \cite{LQGFT}. In particular, the GFT setting should allow for canonical quantization (at least for some, perhaps simplified models). As in the direct canonical quantization of gravitational systems, here one faces the absence of any background time parameter, and hence a problem of time. In the literature one finds two approaches towards defining an operator and Hilbert space formalism of GFT. The first is a more abstract ``timeless'' quantization as proposed in Refs.~\cite{LQGFT,GFTcond}\footnote{To be clear, in this paper we use the term {\em timeless} for a general formalism based on a Hilbert space whose elements do not satisfy dynamical equations, but {\em frozen} for a Dirac-type canonical quantization in which fields initially depend on a parameter $\tau$ but the dynamics state that this dependence is trivial.}: one promotes the GFT field and its conjugate to creation and annihilation operators on a kinematical Hilbert space similar to that of canonical LQG, and then imposes dynamics weakly (in the sense of expectation values, usually in a mean-field approximation). In this approach it is not entirely clear how the choice of original operator algebra is motivated and how the use of unphysical states ({\em i.e.}, states that are not exact solutions to the dynamics) impacts on the validity of the formalism. There is also {\em a priori} no distinction of which operators correspond to observables, although relational observables similar to the canonical quantum gravity setting have been defined \cite{QCGFT}. There is no distinction between kinematical and physical inner product, which is consistent with the fact that the states used are not exact solutions to the dynamics.

In contrast, a ``deparametrized'' canonical quantization for GFT has been studied in Ref.~\cite{edham} following a similar proposal in a GFT toy model \cite{toym}. In this approach one identifies the massless scalar field $\chi$ appearing in some GFT models as a clock variable before quantization and performs the Legendre transform, leading to a conventional quantum theory in which states or observables evolve in $\chi$. This deparametrized formalism can be applied to extract effective cosmological dynamics of GFT, leading to very similar results compared to the timeless formalism \cite{edham,toym,generalcosmology}. Given that the effective cosmology of GFT can be understood from solutions to the classical GFT equations of motion \cite{lowspin}, this agreement is perhaps not surprising, but there are clear differences between the two approaches. The deparametrized approach only works with exact solutions to the dynamics and is hence based on a Hilbert space of physical states (on which there are no further constraints). Some observables which have divergences in the timeless setting are well-behaved in this deparametrized approach. The usual objections of lack of covariance would presumably also apply to the deparametrized quantization in GFT.

By applying the frozen formalism to GFT, we show how a Dirac-type quantization of GFT can be achieved: the timeless Fock space is now interpreted as a kinematical Hilbert space on which constraints are imposed strongly. We again define projections which map operators from the kinematical to a physical Hilbert space, and use these to construct physical observables. We show how the resulting Fock space corresponds to two copies of the Fock space of the deparametrized setting, where the two copies arise since the GFT field is complex whereas it was taken to be real in Ref.~\cite{edham}. Thus, the frozen formalism provides a link between the timeless and deparametrized quantum theories. The effective cosmology obtained in this setting is again analogous to effective Friedmann equations found in previous work starting from Ref.~\cite{QCGFT}. The frozen formalism proposed here may be applicable in more general settings in quantum gravity, since it mimics the basic assumption of Dirac quantization that no time variable should be selected before quantization. In this sense, the frozen formalism should be seen as part of the general program of Dirac (constraint) quantization as an approach to the problem of time; it can bring theories without an obvious constraint structure into a form where a kinematical Hilbert space with subsequent constraint quantization and group averaging may be defined.

In Sec.~\ref{frozenf} we start by reviewing the dynamics of a relativistic particle in Minkowski spacetime and its Dirac quantization as well as the standard canonical quantization of a complex Klein--Gordon field. We then propose a frozen formalism defined in terms of  a new action for the Klein--Gordon field. This action does not change the classical dynamics but it suggests a different route to canonical quantization. We show in which sense this quantization is equivalent to the usual one, and define maps from states and observables in the kinematical to those in the physical Hilbert space. In Sec.~\ref{frozenGFT} we introduce the GFT formalism and review previous proposals for canonical quantization. We then apply the frozen formalism to GFT, where we restrict ourselves to quadratic actions as we do throughout the paper. The dynamics of GFT are similar to Klein--Gordon theory, with the important difference that there are modes with oscillatory solutions but also unstable modes with real exponential solutions. Technical subtleties associated to this property can be overcome by using analytic continuation of the GFT action into the complex $\chi$ plane. In Sec.~\ref{relateff} we show how the frozen formalism leads straightforwardly to the construction of relational observables, which are analogous to those defined previously in the timeless formalism. We also derive a simple effective Friedmann equation, showing that its predictions agree with previous work in GFT cosmology. As an example of an operator on the kinematical Hilbert space that does not become a physical observable, we discuss an operator corresponding to the matter clock itself.

\section{Frozen Formalism for a Klein--Gordon Field}
\label{frozenf}

In this section we introduce a field theoretic version of the frozen formalism appearing in the Dirac quantization of finite-dimensional quantum systems with gauge symmetry under reparametrizations of a ``proper time'' or worldline parameter $\tau$. Given that the usual Klein--Gordon field can be introduced as the many-particle extension of the quantum theory of  a single relativistic particle, this is a natural starting point for our formalism.

\subsection{Relativistic particle}
\label{relparticle}

A relativistic particle in $D$-dimensional Minkowski spacetime is the archetypal example of a dynamical system with reparametrization invariance. It can be defined by a worldline action
\be
S[q^\mu,p_\mu,N]=\int \dd\tau \left(p_\mu\frac{\dd q^\mu}{\dd\tau}+\frac{N}{2}(p^2+m^2)\right)
\label{worldline}
\ee
which is clearly invariant under reparametrizations 
\be
\tau\mapsto\tilde\tau(\tau)\,,\;\frac{\dd q^\mu}{\dd\tau}\mapsto \frac{\dd q^\mu}{\dd\tilde\tau}\,,\; N(\tau)\mapsto \tilde{N}(\tilde\tau)=\frac{N(\tilde\tau)}{\tilde\tau'(\tau)}\,.
\ee
Eq.~(\ref{worldline}) is already in Hamiltonian form: $q^\mu$ and $p_\mu$ are canonically conjugate and the Hamiltonian 
\be
\mathcal{H}=-\frac{N}{2}(p^2+m^2)
\ee
is constrained to vanish by the equation following from varying with respect to $N$. This condition is of course the mass-shell constraint of a relativistic particle.

In canonical (Dirac) quantization \cite{diracqu} one now introduces operators $\hat{q}^\mu$ and $\hat{p}_\nu$ which satisfy
\be
[\hat{q}^\mu,\hat{p}_\nu]=\im\delta^\mu_\nu
\ee
and act on a kinematical Hilbert space $L^2(\mathbb{R}^D)$; for a state $|\psi\rangle$ in this Hilbert space to be considered physical it must satisfy 
\be
\hat{\mathcal{C}}|\psi\rangle =\frac{1}{2}\left(\eta^{\mu\nu}\hat{p}_\mu\hat{p}_\nu + m^2\right)|\psi\rangle = 0\,.
\label{constraint}
\ee
The worldline parameter $\tau$ then disappears from the quantum theory, given that all physical states satisfy
\be
\im\frac{\dd}{\dd\tau}|\psi\rangle = 0
\label{frozeneq}
\ee
since the Hamiltonian $-\hat{N}\hat{\mathcal{C}}$ vanishes when acting on them. The constraint $\hat{\mathcal{C}}$ generates gauge transformations (reparametrizations) in the theory; observables must commute with the constraint to ensure that the action of an observable preserves the space of physical states. Expectation values of observables are then also independent of $\tau$, and one obtains a frozen formalism \cite{frozenrefs}. Dynamical information is encoded in {\em relational} observables, which capture the dynamics of degrees of freedom relative to one another rather than in an external time.

A somewhat subtle point is the definition of a physical inner product. Starting from a basis of states for $L^2(\mathbb{R}^D)$ normalized in the usual improper sense
\be
\langle p|p'\rangle = (2\pi)^D \delta^{(D)}(p-p')\,,
\label{basis}
\ee
one sees that solutions to Eq.~(\ref{constraint}) are not normalizable and hence not elements of this kinematical Hilbert space. The space of physical states hence needs a different inner product which can be constructed by {\em group averaging} \cite{groupav}; writing any physical state as
\be
|\psi_{{\rm ph}}\rangle = \delta(\hat{\mathcal{C}})|\psi\rangle
\ee
where $|\psi\rangle$ is an element of the kinematical Hilbert space, the physical inner product is defined by
\be
\langle \phi_{{\rm ph}}|\psi_{{\rm ph}}\rangle := \langle\phi|\delta(\hat{\mathcal{C}})|\psi\rangle
\label{innerpr}
\ee
where the right-hand side uses the inner product in the kinematical Hilbert space. Concretely, if 
\be
|\psi\rangle = \int \frac{\dd^D p}{(2\pi)^D}\;\psi(p)|p\rangle
\ee
written in terms of the basis (\ref{basis}), one finds\footnote{The additional $\pi$ factor inserted in the constraint is an arbitrary choice made for convenience of normalization.}
\bes
\langle \phi_{{\rm ph}}|\psi_{{\rm ph}}\rangle & = & \int \frac{\dd^D p}{(2\pi)^D} \,\delta\left(\frac{1}{2\pi}(p^2+m^2)\right)\,\overline{\phi(p)}\psi(p)\nonumber
\\&=& \int \frac{\dd^{D-1} p}{(2\pi)^{D-1}} \frac{1}{2\omega_{\vec{p}}}\left(\overline{\phi(\omega_{\vec{p}},\vec{p})}\psi(\omega_{\vec{p}},\vec{p})+\right.\nonumber
\\&&\left.\overline{\phi(-\omega_{\vec{p}},\vec{p})}\psi(-\omega_{\vec{p}},\vec{p})\right)
\label{physinner}
\ees
where $\omega_{\vec{p}}:=\sqrt{\vec{p}^2+m^2}$ (see, {\em e.g.}, Ref.~\cite{switch} for more details and discussion). Eq.~(\ref{physinner}) is equivalent to the standard relativistic inner product for solutions to the Klein--Gordon equation, with a sign flipped to make it positive for positive and negative frequency states (for which $p_0=\pm \omega_{\vec{p}}$, respectively). Although it can be represented as an integral over spatial momenta only, Eq.~(\ref{innerpr}) shows that this inner product is Lorentz invariant. Maintaining the symmetries of the Lagrangian theory is one argument for constructing an inner product via group averaging.

Eq.~(\ref{physinner}) then shows that the physical Hilbert space is $L^2(\mathbb{R}^{D-1})_+ \oplus L^2(\mathbb{R}^{D-1})_-$, the direct sum of two Hilbert spaces for positive and negative frequency states. Writing a physical state as
\be
|\psi_{{\rm ph}}\rangle = \int \frac{\dd^{D-1} p}{(2\pi)^{D-1}}\frac{1}{\sqrt{2\omega_{\vec{p}}}}\left(\psi^+(\vec{p})|\vec{p},+\rangle+\psi^-(\vec{p})|\vec{p},-\rangle\right)
\label{newdecomp}
\ee
defines a new basis for this physical Hilbert space with normalization
\be
\langle \vec{p},\pm|\vec{p}',\pm'\rangle = (2\pi)^{D-1} \delta_{\pm,\pm'}\delta^{(D-1)}(\vec{p}-\vec{p}')\,.
\label{newbasis}
\ee
The wavefunctions $\psi^{\pm}$ in Eq.~(\ref{newdecomp}) correspond to the components in Eq.~(\ref{physinner}) for which $p_0=\pm \omega_{\vec{p}}$.

\subsection{Conventional Klein--Gordon theory}

The previous constructions define a consistent quantum theory of a single relativistic particle. ``Second quantization'' of this theory leads to a quantum field theory for a complex scalar field $\Phi$. Here the constraint (\ref{constraint}) is not imposed as an equation on the one-particle Hilbert space but becomes the equation of motion for $\Phi$,
\be
\left(\eta^{\mu\nu}p_\mu p_\nu + m^2\right)\Phi(p)=0\,.
\ee
This field equation can be derived from an action 
\be
S[\Phi,\overline\Phi] = -\int \frac{\dd^D p}{(2\pi)^D}\, \overline{\Phi}(p) \left(\eta^{\mu\nu}p_\mu p_\nu + m^2\right)\Phi(p)
\label{kgaction}
\ee
which can then again be used as a starting point for canonical quantization. Conjugate momenta to the field variables are obtained after Fourier transform from $p^0$ to a time coordinate $t$,
\be
\pi(t,\vec{p})=\frac{\partial\mathcal{L}}{\partial (\partial_t\Phi(t,\vec{p}))}\,,\quad\overline\pi(t,\vec{p})=\frac{\partial\mathcal{L}}{\partial( \partial_t{\overline{\Phi}}(t,\vec{p}))}\,,
\ee
and one can rewrite the action as
\bes
S &=& \int \dd t\,\frac{\dd^{D-1} p}{(2\pi)^{D-1}}\left(\pi\partial_t\Phi+\overline\pi\partial_t\overline\Phi-\mathcal{H}\right)\,,\nonumber
\\\mathcal{H} & = & |\pi|^2 + (\vec{p}^2+m^2)|\Phi|^2\,.
\ees

The canonical variables are then promoted to operators satisfying
\bes
[\hat\Phi(t,\vec{p}),\hat\pi(t,\vec{p}')] &=& [ {\hat{\Phi}}^\dagger(t,\vec{p}),{\hat{\pi}}^\dagger(t,\vec{p}') ] \nonumber
\\ &=& (2\pi)^{D-1}\im\delta^{(D-1)}(\vec{p}-\vec{p}')\,.
\ees
To diagonalize the Hamiltonian one can introduce two sets of annihilation operators
\bes
\hat{a}(\vec{p})&=&\frac{1}{\sqrt{2\omega_{\vec{p}}}}\left(\omega_{\vec{p}}\hat\Phi(\vec{p})+\im{\hat{\pi}}^\dagger(\vec{p})\right)\,,\nonumber
\\\hat{b}(\vec{p})&=&\frac{1}{\sqrt{2\omega_{\vec{p}}}}\left(\omega_{\vec{p}}\hat\Phi^\dagger(\vec{p})+\im{\pi}(\vec{p})\right)\,,
\label{ladderop}
\ees
with their Hermitian conjugates acting as creation operators; writing the Hamiltonian as $\hat{\mathcal{H}}=\int \frac{\dd^{D-1} p}{(2\pi)^{D-1}}\hat{\mathcal{H}}(\vec{p})$ one then finds (after normal ordering)
\be
\hat{\mathcal{H}}(\vec{p})=\omega_{\vec{p}}\left(\hat{a}^\dagger(p)\hat{a}(\vec{p})+\hat{b}^\dagger(p)\hat{b}(\vec{p})\right)\,.
\label{modehamiltonian}
\ee
One can think of the creation and annihilation operators as time-dependent, inheriting the $t$ dependence of the dynamical fields, or construct them from fields at some initial time $t=0$. The two sets of creation operators $\hat{a}^\dagger$ and $\hat{b}^\dagger$ are associated with particles and antiparticles. We will adopt the convention that these operators do not evolve in time.

Eq.~(\ref{ladderop}) means that the time-dependent original fields can be written as
\bes
\hat\Phi(t,\vec{p})&=&\frac{1}{\sqrt{2\omega_{\vec{p}}}}\left(e^{-\im\omega_{\vec{p}}t}\hat{a}(\vec{p})+e^{\im\omega_{\vec{p}}t}\hat{b}^\dagger(\vec{p})\right)\,,\nonumber
\\ \hat\Phi^\dagger(t,\vec{p})&=&\frac{1}{\sqrt{2\omega_{\vec{p}}}}\left(e^{\im\omega_{\vec{p}}t}\hat{a}^\dagger(\vec{p})+e^{-\im\omega_{\vec{p}}t}\hat{b}(\vec{p})\right)
\label{timedepf}
\ees
where we have written out the time dependence on the right-hand side explicitly.
A one-particle state at $t=0$ can then be written as
\be
|\psi\rangle = \int \frac{\dd^{D-1} p}{(2\pi)^{D-1}} \frac{1}{\sqrt{2\omega_{\vec{p}}}}\left(\psi^+(\vec{p})\hat{a}^\dagger(\vec{p})+\psi^-(\vec{p})\hat{b}^\dagger(\vec{p})\right)|0\rangle 
\ee
where $|0\rangle$ is the Fock vacuum annihilated by all annihilation operators; the inner product between two such one-particle states is
\be
\langle \phi|\psi \rangle = \int \frac{\dd^{D-1} p}{(2\pi)^{D-1}} \frac{1}{2\omega_{\vec{p}}}\left(\overline{\phi^+(\vec{p})}\psi^+(\vec{p}) + \overline{\phi^-(\vec{p})}\psi^-(\vec{p})\right)
\label{1pinnerproduct}
\ee
which is the same as the inner product (\ref{physinner}). The one-particle sector of the Fock space is exactly the physical Hilbert space of the relativistic particle, {\em i.e.}, the pairs of functions $(\phi^+,\phi^-)$ and $(\psi^+,\psi^-)$ in Eq.~(\ref{1pinnerproduct}) are again elements of $L^2(\mathbb{R}^{D-1})_+ \oplus L^2(\mathbb{R}^{D-1})_-$. For each $\vec{p}$ the two solutions to the constraint $p^2+m^2=0$ are now associated with particle and antiparticle excitations. The states defined in Eq.~(\ref{newbasis}) can be identified with
\be
|\vec{p},+\rangle = \hat{a}^\dagger(\vec{p})|0\rangle\,,\quad |\vec{p},-\rangle = \hat{b}^\dagger(\vec{p})|0\rangle\,.
\label{identif}
\ee

\subsection{Frozen formalism}
\label{frozensec}

The usual canonical quantization of a complex scalar field extends the physical Hilbert space of a relativistic particle constructed in Section \ref{relparticle} to a Fock space which contains arbitrary numbers of such particles. However, this viewpoint on quantum field theory shows no trace of the Dirac quantization performed to construct the physical Hilbert space of a relativistic particle; there is no reparametrization invariance in the Klein--Gordon theory, no kinematical Hilbert space, and no analog of the condition (\ref{frozeneq}). In this subsection we propose a quantization of the complex scalar field which has these features, and thus provides a field theory extension of the frozen formalism of Dirac quantization.

Recall that the Schr\"odinger equation $\im\frac{\partial \psi}{\partial t}=\hat{\mathcal{H}}\psi$ can be derived from the action
\be
S[\psi,\overline\psi]=\int \dd X\,\dd t\,\left[\frac{\im}{2}\left(\overline\psi\frac{\partial\psi}{\partial t}-\psi\frac{\partial\overline\psi}{\partial t}\right)-\overline\psi\hat{\mathcal{H}}\psi\right]
\label{schract}
\ee
where $X$ denotes the configuration space of the theory one is studying, and $\hat{\mathcal{H}}$ is a differential operator acting on the $X$ variables which becomes the Hamiltonian in the quantum theory. The quantum theory of the relativistic particle can be seen as defined by a Schr\"odinger equation for which only zero-energy states are allowed. This motivates the definition of the field theory action
\bes
S[\Phi,\overline\Phi,N]&=&\int \frac{\dd^D p}{(2\pi)^D}\,\dd \tau\Big[\frac{\im}{2}\left(\overline\Phi\frac{\partial\Phi}{\partial \tau}-\Phi\frac{\partial\overline\Phi}{\partial \tau}\right)\nonumber
\\&&+N(p^2+m^2)|\Phi|^2\Big]\,.
\label{newaction}
\ees
Notice the similarity of Eq.~(\ref{newaction}) with the worldline action (\ref{worldline}) for the relativistic particle: the fields $\Phi$ and $\overline\Phi$ now depend on a parameter $\tau$. There is a significant literature on such formulations of relativistic quantum mechanics in which one introduces a ``worldline'' or ``proper time'' parameter $\tau$ \cite{review}. For instance, in a theory in which the Hamiltonian is not constrained to vanish, different energy eigenstates correspond to all possible values for the squared mass $m^2$ which is then no longer a fundamental parameter of the theory. Here we will follow a Dirac quantization, require constrained dynamics and allow only zero-energy states, similar to the discussion of, {\em e.g.}, Ref.~\cite{nambu}. The equations of motion following from Eq.~(\ref{newaction}) are
\bes
\im\frac{\partial\Phi}{\partial \tau}+N(p^2+m^2)\Phi&=&0\,,\nonumber
\\-\im\frac{\partial\overline\Phi}{\partial\tau}+N(p^2+m^2)\overline\Phi&=&0\,,\nonumber
\\(p^2+m^2)|\Phi|^2&=&0\,.
\ees
The last equation implies that $\Phi(\tau,p)=\overline\Phi(\tau,p)=0$ unless $p^2+m^2=0$ and the first two equations then say that $\Phi$ and $\overline\Phi$ must be independent of $\tau$. The theory has a reparametrization invariance which is rather trivial. Classically, the theory defined by Eq.~(\ref{newaction}) is equivalent to the one defined by Eq.~(\ref{kgaction}).

One can now proceed with Dirac quantization. Eq.~(\ref{newaction}) implies that $\Phi$ and $\im\overline\Phi$ are canonically conjugate; the corresponding operators satisfy
\be
[\hat\Phi(p),\hat\Phi^\dagger(p')]=(2\pi)^D \delta^{(D)}(p-p')
\label{canonconj}
\ee
as they do in usual non-relativistic quantum field theory based on actions of the form (\ref{schract}), but {\em unlike} in the standard quantization of a relativistic field theory in which the field operators commute. The field operators defined by Eq.~(\ref{canonconj}) act as creation and annihilation operators on a {\em kinematical} Hilbert space. If we apply normal ordering, physical states must satisfy the constraint(s)
\be
\hat{\mathcal{C}}(p):=(p^2+m^2)\hat\Phi^\dagger(p)\hat\Phi(p)|\psi\rangle = 0\,.
\label{fockconst}
\ee
$\hat\Phi^\dagger(p)\hat\Phi(p)=:\hat{N}(p)$ is the number density operator for the mode $p$; physical states are those for which only physical modes with $p^2+m^2=0$ are excited. To construct a physical Hilbert space for solutions of Eq.~(\ref{fockconst}), we follow the same steps as in Section \ref{relparticle} for one-particle states 
\be
|\psi\rangle = \int \frac{\dd^D p}{(2\pi)^D}\;\psi(p)\,\hat\Phi^\dagger(p)|0\rangle_{{\rm kin}}
\ee
where $|0\rangle_{{\rm kin}}$ is the Fock vacuum of the kinematical Hilbert space; for such states, in analogy with Eq.~(\ref{physinner}) we define 
\bes
\langle \phi_{{\rm ph}}|\psi_{{\rm ph}}\rangle & = & \int \frac{\dd^D p}{(2\pi)^D} \,\delta\left(\frac{1}{2\pi}(p^2+m^2)\right)\,\overline{\phi(p)}\psi(p)\nonumber
\\&=& \int \frac{\dd^{D-1} p}{(2\pi)^{D-1}} \frac{1}{2\omega_{\vec{p}}}\left(\overline{\phi(\omega_{\vec{p}},\vec{p})}\psi(\omega_{\vec{p}},\vec{p})+\right.\nonumber
\\&&\left.\overline{\phi(-\omega_{\vec{p}},\vec{p})}\psi(-\omega_{\vec{p}},\vec{p})\right)
\label{physinner2}
\ees
where $|\psi_{{\rm ph}}\rangle = \delta(p^2+m^2)|\psi\rangle$ for a general one-particle state $|\psi\rangle$. In general a formal insertion of $\delta(\hat{\mathcal{C}}(p))$ into the inner product would lead to the question of how to define $\delta(\hat{N}(p))$; here, instead of trying to make such a definition more rigorous, we first only define the inner product for single-particle states through Eq.~(\ref{physinner2}). This construction then leads to exactly the physical Hilbert space of single particles or antiparticles defined in the previous sections, {\em i.e.}, the Hilbert space $L^2(\mathbb{R}^{D-1})_+ \oplus L^2(\mathbb{R}^{D-1})_-$.

We can extend the construction of a physical inner product to the entire kinematical Fock space generated by the repeated action of $\hat\Phi^\dagger(p)$ on $|0\rangle_{{\rm kin}}$: this extension is determined by the requirement that the physical Hilbert space is also a Fock space, the ``second quantization'' of the physical one-particle Hilbert space. 

Just as for the Dirac quantization of the relativistic particle, this physical Fock space is not a subspace of the kinematical Fock space, but is obtained by the action of a different set of creation operators on a different vacuum which we denote by $|0\rangle_{{\rm ph}}$. In order to make the relation between the two separate Hilbert spaces explicit we define a ``projection'' $\mathcal{P}$ (clearly not a projection in the usual sense) of creation and annihilation operators 
\be
\hat\Phi(p)\mapsto \mathcal{P}\hat\Phi(p)\,,\quad \hat\Phi^\dagger(p)\mapsto \mathcal{P}\hat\Phi^\dagger(p)\,,
\label{Pmap}
\ee
such that the projected operators generate a physical Fock space when acting on $|0\rangle_{{\rm ph}}$. In order to fix the explicit form of $\mathcal{P}$, we now demand that the inner product for the one-particle sector has to be consistent with Eq.~(\ref{physinner2}). This implies that we need
\be
\mathcal{P}\hat\Phi^\dagger(p)=\frac{2\pi}{\sqrt{2\omega_{\vec{p}}}}\left(\delta(p^0-\omega_{\vec{p}})\hat{a}^\dagger(\vec{p})+\delta(p^0+\omega_{\vec{p}})\hat{b}^\dagger(\vec{p})\right)
\label{Pnormal}
\ee
with an analogous definition (obtained by Hermitian conjugate) for $\mathcal{P}\hat\Phi(p)$ and with canonical commutators
\be
[\hat{a}(\vec{p}),\hat{a}^\dagger(\vec{p}')]  =  [\hat{b}(\vec{p}),\hat{b}^\dagger(\vec{p}')]  =  (2\pi)^{D-1}\delta^{(D-1)}(\vec{p}-\vec{p}')
\ee
for the newly introduced creation and annihilation operators acting on $|0\rangle_{{\rm ph}}$.
These definitions imply a map
\bes
|\psi\rangle &=& \int \frac{\dd^D p}{(2\pi)^D}\psi(p)\hat\Phi^\dagger(p)|0\rangle_{{\rm kin}}\nonumber
\\\mapsto\; |\psi_{{\rm ph}}\rangle &=& \int \frac{\dd^D p}{(2\pi)^D}\psi(p)\mathcal{P}\hat\Phi^\dagger(p)|0\rangle_{{\rm ph}}
\label{statemap}
\\&=& \int \frac{\dd^{D-1} p}{(2\pi)^{D-1}}\frac{1}{\sqrt{2\omega_{\vec{p}}}}\left(\psi(\omega_{\vec{p}},\vec{p})\hat{a}^\dagger(\vec{p})+\right.\nonumber
\\&&\left.\quad\psi(-\omega_{\vec{p}},\vec{p})\hat{b}^\dagger(\vec{p})\right)|0\rangle_{{\rm ph}}\nonumber
\ees
from the kinematical to the physical one-particle Hilbert space; the map then extends to arbitrary Fock states by writing these as the result of the action of some operator on the vacuum $|0\rangle_{{\rm kin}}$ and then applying the map (\ref{Pmap}). For instance, the physical inner product for two-particle states becomes (schematically) 
\be
\langle \phi_{{\rm ph}}|\psi_{{\rm ph}}\rangle=\langle\phi|\delta\left(\frac{1}{2\pi}(\hat{p}_1^2+m^2)\right)\delta\left(\frac{1}{2\pi}(\hat{p}_2^2+m^2)\right)|\psi\rangle
\ee
which involves two insertions of a constraint corresponding to the momenta of the two particles involved. This generalized notion of Dirac quantization of a complex scalar field based on the action (\ref{newaction}) is then equivalent to the canonical quantization based on Eq.~(\ref{kgaction}).

One can define similar projection maps from Dirac observables on the kinematical Hilbert space\footnote{These are Hermitian operators that commute ``weakly'' with the constraints, {\em i.e.}, commutators with constraints vanish if Eq.~(\ref{fockconst}) holds. If these commutators are non-zero, self-adjointness on the physical Hilbert space is a non-trivial requirement \cite{switch}.} to observables on the physical Hilbert space. For instance, consider the particle number 
\be
\hat{N} = \int \frac{\dd^D p}{(2\pi)^D}\hat\Phi^\dagger(p)\hat\Phi(p)
\label{kinenumb}
\ee
which is a Dirac observable with discrete spectrum $\mathbb{N}_0$ on the kinematical Fock space. The number operator on the physical Fock space is
\be
\hat{N}_{{\rm ph}} = \int \frac{\dd^{D-1} p}{(2\pi)^{D-1}}\left(\hat{a}^\dagger(\vec{p})\hat{a}(\vec{p})+\hat{b}^\dagger(\vec{p})\hat{b}(\vec{p})\right)\,.
\label{physnumb}
\ee
If we demand that the projection map acting on bilinear field operators maps the number operator on the kinematical Fock space to the one on the physical Fock space, this fixes this map to be
\bes
\hat\Phi^\dagger(p)\hat\Phi(p)&\mapsto& \mathcal{P}\left(\hat\Phi^\dagger(p)\hat\Phi(p)\right)
\label{fieldproj}
\\&=& 2\pi\left(\delta(p^0-\omega_{\vec{p}})\hat{a}^\dagger(\vec{p})\hat{a}(\vec{p})+\delta(p^0+\omega_{\vec{p}})\hat{b}^\dagger(\vec{p})\hat{b}(\vec{p})\right)\,.\nonumber
\ees
This map differs from the one defined for single field operators in Eq.~(\ref{Pnormal}). First of all we chose a different normalization, fixed by the requirement that the number operator on the physical Hilbert space has spectrum $\mathbb{N}_0$ ({\em i.e.}, particles are counted in units of 1). In contrast, a change in normalization in Eq.~(\ref{Pnormal}) could be absorbed in a redefinition of the wavefunction in Eq.~(\ref{statemap}). This difference is therefore more a matter of conventions. More importantly however, one cannot simply apply Eq.~(\ref{Pnormal}) separately to each field operator in Eq.~(\ref{kinenumb}), since this would result in an additional factor $\delta(0)$.  The reason for this is again the difference between the kinematical and physical inner product. Similar considerations will apply in the rest of the paper: ``projection'' maps always have to be defined separately for single field operators and composite operators.

Eq.~(\ref{fieldproj}) can be applied to more general Dirac observables of the form 
\be
\hat{O}_f = \int \frac{\dd^D p}{(2\pi)^D}f(p)\hat\Phi^\dagger(p)\hat\Phi(p)
\label{genobs}
\ee
on the kinematical Hilbert space. One example is the energy $\hat{E}=\hat{O}_{|p^0|}$ (absolute values ensure that all excitations are associated with a positive energy), which after applying Eq.~(\ref{fieldproj}) becomes
\be
\hat{E}_{{\rm ph}} = \int \frac{\dd^{D-1} p}{(2\pi)^{D-1}}\,\omega_{\vec{p}}\left(\hat{a}^\dagger(\vec{p})\hat{a}(\vec{p})+\hat{b}^\dagger(\vec{p})\hat{b}(\vec{p})\right)
\ee
which is the usual Hamiltonian (\ref{modehamiltonian}). $\hat{E'}=\hat{O}_{p^0}$ is an equally well-defined observable, which associates negative energy to antiparticle excitations. 

Creation operators constructed from Eq.~(\ref{Pnormal})  can be seen as defining a physical Hilbert space in the Heisenberg picture: states are time-independent, arising from action of creation operators at $t=0$ on a Fock vacuum. Observables of the form (\ref{genobs}) are time-independent in either Schr\"odinger or Heisenberg picture, since all particle number densities are conserved in the theory (for particle and antiparticle sector separately). One might alternatively be interested in time-dependent (Schr\"odinger) Fock states or equivalently time-dependent single field operators, which can be used to create time-dependent states by acting on the vacuum $|0\rangle_{{\rm ph}}$. Such operators can be obtained from Eq.~(\ref{Pnormal}) by inserting a time evolution factor $e^{\im p^0 t}$, which yields a time-dependent projection
\bes
\mathcal{P}_t\hat\Phi^\dagger(p)&=&\frac{2\pi}{\sqrt{2\omega_{\vec{p}}}}\left(e^{\im\omega_{\vec{p}}t}\delta(p^0-\omega_{\vec{p}})\hat{a}^\dagger(\vec{p})\right.\nonumber
\\&&\left.+e^{-\im\omega_{\vec{p}}t}\delta(p^0+\omega_{\vec{p}})\hat{b}^\dagger(\vec{p})\right)\,.
\label{timedepproj}
\ees
Looking at the time dependence of a physical Fock state defined in this way, one sees that particle and antiparticle states evolve with opposite phase factors, in contrast to the usual quantization where they both evolve as $e^{-\im\omega_{\vec{p}}t}$. The reason for this difference is the same we saw in defining the energy; the kinematical variable $p^0$ is positive for the particle sector but negative for the antiparticle sector. This unusual property can be traced back to the difference between Eq.~(\ref{timedepproj}) and the relation (\ref{timedepf}) in conventional Klein--Gordon theory and fundamentally to the fact that the field $\hat\Phi$ and its Hermitian conjugate $\hat\Phi^\dagger$ defined by Eq.~(\ref{timedepf}) commute whereas here they do not.

Particle and antiparticle sectors are decoupled and there is no operational way of distinguishing the sign of a phase factor, so this issue is not relevant for the physical content of the theory, but it would become relevant and potentially problematic for an interacting theory (which we will not study in this paper). 

It is possible to define time-dependent observables on the physical Hilbert space as well. Consider for instance
\be
\hat{Q}_f = \int \frac{\dd^D p}{(2\pi)^D}f(p)\hat\Phi^\dagger(p)\hat\Phi(-p)
\label{weirdob}
\ee
with $f(p)=\overline{f(-p)}$. This operator commutes weakly with the constraints (\ref{fockconst}): the commutator of $\hat{Q}_f$ with a constraint $ \hat{\mathcal{C}}(p)$ vanishes on all physical states. Hence this operator preserves the space of physical states defined by Eq.~(\ref{fockconst}), and defines a Dirac observable on the kinematical Hilbert space (cf.~the discussion above Eq.~(\ref{kinenumb})).

Let us first consider the projection of such an observable to an observable on the physical Fock space at time $t=0$, when Heisenberg and Schr\"odinger pictures agree. Instead of Eq.~(\ref{fieldproj}) we now define
\bes
\mathcal{P}\left(\hat\Phi^\dagger(p)\hat\Phi(-p)\right)&=&2\pi\left(\delta(p^0-\omega_{\vec{p}})\hat{a}^\dagger(\vec{p})\hat{b}(-\vec{p})+\right.\nonumber
\\&&\left.\delta(p^0+\omega_{\vec{p}})\hat{b}^\dagger(\vec{p})\hat{a}(-\vec{p})\right)\qquad
\ees
using again the normalization derived from the number operator (\ref{physnumb}) and making sure the resulting operator is well-defined in the inner product on the physical Fock space. Away from $t=0$, we can either work in the Schr\"odinger picture where states evolve according to Eq.~(\ref{timedepproj}), or in the Heisenberg picture where observables such as $\mathcal{P}\hat{Q}_f$ should instead evolve in time. As usual, demanding that expectation values agree in the two pictures fixes the time dependence of operators in the Heisenberg picture; here we find that  we need to extend the time-dependent map (\ref{timedepproj}) to composite operators by
\bes
\mathcal{P}_t\left(\hat\Phi^\dagger(p)\hat\Phi(-p)\right)&=&2\pi\left(e^{-2\im\omega_{\vec{p}}t}\delta(p^0-\omega_{\vec{p}})\hat{a}^\dagger(\vec{p})\hat{b}(-\vec{p})+\right.\nonumber
\\&&\left.e^{2\im\omega_{\vec{p}}t}\delta(p^0+\omega_{\vec{p}})\hat{b}^\dagger(\vec{p})\hat{a}(-\vec{p})\right)\qquad
\ees
and hence
\bes
\mathcal{P}_t\hat{Q}_f &=& \int\frac{\dd^{D-1} p}{(2\pi)^{D-1}}\left(f(\omega_{\vec{p}},\vec{p})\,e^{-2\im\omega_{\vec{p}}t}\,\hat{a}^\dagger(\vec{p})\hat{b}(-\vec{p})+\right.\nonumber
\\&&\left.\overline{f(\omega_{\vec{p}},\vec{p})}\,e^{2\im\omega_{\vec{p}}t}\,\hat{b}^\dagger(-\vec{p})\hat{a}(\vec{p})\right)
\ees
which is a Hermitian operator on the physical Fock space defined in the Heisenberg picture. The time dependence of this observable is due to a mixing of particle and antiparticle sectors, and its physical interpretation may be unclear at this point.

\section{Frozen formalism in group field theory}
\label{frozenGFT}

Background-independent approaches to quantum gravity encounter a problem of time due to the absence of a global background time parameter. This is particularly apparent in approaches to canonical quantization of gravitational systems such as in quantum cosmology or LQG, where the methods of Dirac quantization have mostly been applied \cite{groupav}. However, there are other approaches in which one does not directly quantize the degrees of freedom of classical gravity, but expects gravitational dynamics to emerge from the interaction of different (``non-spatiotemporal'') quantum degrees of freedom. Such approaches also face a problem of time if they are to be compatible with general covariance. Here we will focus on group field theory (GFT) \cite{GFT} which incorporates much of the structure of canonical LQG while also being formulated in the language of  quantum field theory. We can use the insights from our discussion of Klein--Gordon theory to define a frozen canonical quantization of GFT.

\subsection{Group field theory formalism and canonical quantization}
\label{GFTintro}

In the GFT models we consider, the basic variable is a complex scalar field $\varphi$ whose arguments are elements of a Lie group, here taken to be four copies of ${\rm SU}(2)$, and a real-valued (scalar) matter field variable $\chi$. The field is ``gauge invariant'' with respect to its ${\rm SU}(2)$ arguments,
\be
\varphi(g_1,\ldots,g_4,\chi)=\varphi(g_1h,\ldots,g_4h,\chi)\quad\forall\;h\in{\rm SU}(2)\,,
\label{gaugeinv}
\ee
and its dynamics are defined in terms of an action
\be
S[\varphi,\bar\varphi]=\int \dd^4 g\;\dd\chi\;\bar\varphi(g_I,\chi)\mathcal{K}\varphi(g_I,\chi) + V[\varphi,\bar\varphi]
\label{gftaction}
\ee
where $V[\varphi,\bar\varphi]$ includes the interaction terms which are usually of fourth and higher order in the fields. The kernel $\mathcal{K}$ in Eq.~(\ref{gftaction}) can in general be a nonlocal operator acting on $\varphi$ but we will assume that $\mathcal{K}$ can be written in terms of a finite number of derivatives and coupling constants. For GFT models for quantum gravity, $\mathcal{K}$ is often taken to be initially trivial, {\em i.e.}, just a constant \cite{SFGFT}, but radiative corrections then generate Laplace--Beltrami derivative operators with respect to the arguments of the field \cite{radcorr}. Within a more general class of models in which $\mathcal{K}$ is nonlocal, one could obtain a local form from considering the first few terms in a derivative expansion \cite{QCGFT}.

The connection of GFT to quantum gravity is made by expanding the GFT partition function perturbatively around the free theory,
\be
Z = \int \mathcal{D}\varphi\;\mathcal{D}\bar\varphi\;e^{-S[\varphi,\bar\varphi]} = \sum_\Gamma \frac{\lambda^{V(\Gamma)}}{{\rm sym}(\Gamma)}\;A[\Gamma]
\label{gftexpansion}
\ee
where we have for simplicity assumed a single interaction term including a coupling $\lambda$. The sum in Eq.~(\ref{gftexpansion}) is over Feynman graphs $\Gamma$; $V(\Gamma)$ is the number of vertices in $\Gamma$ and ${\rm sym}(\Gamma)$ a symmetry factor.  For a suitable definition of the interaction term, one can then identify each $\Gamma$ with a discrete spacetime history and $A[\Gamma]$ with a {\em spin foam} amplitude associated to $\Gamma$, {\em i.e.}, with a discrete quantum gravity (or topological field theory) path integral including a sum over all geometric data on $\Gamma$ \cite{SFGFT}. Hence, assuming one can somehow make mathematical sense of it, the GFT partition function generates a sum over all possible discrete spacetime histories weighted by quantum gravity amplitudes, and can be argued to define a proposal for a theory of quantum gravity. This correspondence is particularly well understood for topological models such as the Ooguri model \cite{ooguri} for which the amplitudes are those of a topological field theory.

We are interested in defining a canonical quantization of Eq.~(\ref{gftaction}). For simplicity we will only consider the free theory. Interactions contained in $V[\varphi,\bar\varphi]$ can be included perturbatively as is standard in canonical quantization, although we leave this to future work. The first step is then to bring Eq.~(\ref{gftaction}) into a simpler form by using the Peter--Weyl decomposition of functions on ${\rm SU}(2)$ into irreducible representations. Define
\be
\varphi(g_I,\chi)= \sum_J \varphi_{J}(\chi) D^J(g_I)\,,\;\bar\varphi(g_I,\chi) = \sum_J \bar\varphi_{J}(\chi) \overline{D^J(g_I)}
\label{modedecomp}
\ee
where $J=(j_I,m_I,\iota)$ is a multi-index\footnote{This economical notation was introduced in Ref.~\cite{mehdiisha}.} depending on four irreducible representations $j_I\in\mathbb{N}_0/2$, magnetic indices $m_I\in\{-j_I,-j_I+1,\ldots,+j_I\}$ and intertwiners $\iota$ (${\rm SU}(2)$ invariant maps from the tensor product $\otimes j_I$ to the trivial representation). $D^J(g_I)$ are convolutions of Wigner $D$-matrices defined by
\be
D^J(g_I)=\sum_{n_I} \mathcal{I}_{n_I}^{j_I,\iota} \prod_{K=1}^4 \sqrt{2j_K+1}\,{D^{j_K}(g_K)}^{m_K}_{n_K}
\ee
where $D^{j}(g)$ are the usual Wigner matrices for the representation $j$, $\mathcal{I}_{n_I}^{j_I,\iota}$ is the intertwiner for $j_I$ labeled by $\iota$ and the normalization has been fixed so that
\be
\int \dd^4 g\; \overline{D^J(g_I)}\,D^{J'}(g_I)=\delta_{J,J'}=\delta_{j_I,j'_I}\delta_{m_I,m'_I}\delta_{\iota,\iota'}.
\ee
The action (\ref{gftaction}), now restricted to its free part $S_{{\rm f}}$ and with $\chi$ derivatives truncated at second order as in \cite{QCGFT}, then takes the form
\be
S_{{\rm f}}[\varphi,\bar\varphi]=\sum_J\int \dd\chi\;\bar\varphi_J(\chi)\left(\mathcal{K}_J^{(0)}+\mathcal{K}_J^{(2)}\partial_\chi^2\right)\varphi_J(\chi)
\label{freeakshn}
\ee
which is our starting point for canonical quantization. In the literature one finds two approaches to defining a Hilbert space quantization of Eq.~(\ref{freeakshn}). One is based on identifying the matter variable $\chi$ with time (before quantization) and performing a standard Legendre transform, which results in a Fock space built from creation and annihilation operators on which a conventional Hamiltonian evolution is defined, as in the canonical quantization of usual bosonic quantum field theory \cite{edham}. From the perspective of quantum gravity, this strategy is analogous to {\em deparametrization}, in which a time variable is identified among the dynamical degrees of freedom before quantization \cite{deparam}. Indeed a (free) massless scalar is often used as a clock in deparametrization in canonical quantum gravity, which was a main motivation for introducing it also into GFT \cite{QCGFT}. 

As we noted before, deparametrization amounts to a gauge fixing before quantization; the resulting theory is not generally covariant and one would need to show later that the resulting theory does not depend on the choice of clock. In general, no obvious candidate for a global clock may be available. Dirac quantization can define a more covariant notion of quantization. 

A different quantization for GFT (which we here call {\em timeless}) was introduced in Refs.~\cite{GFTcond,LQGFT}. Here the classical fields are promoted to operators satisfying
\be
[\hat\varphi_J(\chi),\hat\varphi^\dagger_{J'}(\chi')]=\delta_{J,J'}\delta(\chi-\chi')\,.
\label{timelessccr}
\ee
These field operators can be seen as creation and annihilation operators on a Fock space, where they generate quanta labeled by representation labels $J$ and matter field values $\chi$, such that states with different $J$ or $\chi$ labels are orthogonal. This is precisely the structure of (kinematical) LQG states on a graph formed by four links meeting at a vertex\footnote{This is the sense in which Eq.~(\ref{gaugeinv}) ensures gauge invariance: with respect to ${\rm SU}(2)$ gauge transformations at a vertex in LQG.} if we extend the ${\rm SU}(2)$ holonomy variables of LQG by a real-valued matter field at the vertex, in slight generalization of canonical LQG where this matter field would be valued in ${\rm U}(1)$ \cite{qsd}. Repeated action of creation operators on the Fock vacuum corresponds to adding more vertices and links to the graph, and by integrating over common arguments one can generate states that correspond to LQG states on arbitrary four-valent graphs \cite{LQGFT}. This correspondence between Fock states generated by Eq.~(\ref{timelessccr}) and quantum states in canonical LQG was one of the main motivations for this timeless quantization. Since the GFT dynamics have not been used to obtain Eq.~(\ref{timelessccr}), this structure is purely kinematical (just as the LQG Hilbert space in relation to the Hamiltonian constraint). 

The philosophy starting from Ref.~\cite{GFTcond} has been to use Eq.~(\ref{timelessccr}) to define a Hilbert space on which dynamics are imposed, {\em e.g.}, by demanding that the GFT equations of motion are satisfied in expectation values for a class of coherent states, leading to a {\em mean-field approximation} in which one solves the classical GFT equations of motion. This approximation is the basis for many results in the application to cosmology \cite{QCGFT}. The fact that one assumed Eq.~(\ref{timelessccr}) then often appears inconsequential, given that one only deals with classical field equations. Once the formalism is pushed further, unusual features appear: two-point functions for cosmological observables such as volume fluctuations, evaluated in the inner product induced by Eq.~(\ref{timelessccr}), are formally singular and require regularization \cite{GFTpert,mehdiisha2}. Moreover, formally any Hermitian operator becomes an observable on the GFT Fock space, which allows the definition of, {\em e.g.}, a ``total scalar field operator'' which sums up all $\chi$ labels in a general state. The connection of this operator to the interpretation of $\chi$ as a relational clock is not clear (see however Ref.~\cite{lucadaniele} for a possible effective relational interpretation). This is in contrast with the deparametrized quantization in which states or operators evolve in $\chi$, which becomes a label as in usual quantum mechanics.

\subsection{Frozen GFT}
\label{frozengftsec}

The viewpoint we want to adopt in this paper is that Eq.~(\ref{timelessccr}) should be understood as defining field operators on a {\em kinematical} Hilbert space in the sense of Dirac quantization, {\em i.e.}, a Hilbert space whose states are not physical unless they satisfy constraints. In contrast to previous work on the timeless approach in which constraints are imposed weakly, here we advocate a strong imposition of constraints. We argue that doing this clarifies the link between deparametrized and timeless approaches.

As in the case of Klein--Gordon theory it is best to work in momentum space, where constraints become decoupled equations for each mode in the kinematical Fock space. Our starting point is still the free GFT action
\be
S_{{\rm f}}[\varphi,\bar\varphi]=\sum_J\int \dd\chi\;\bar\varphi_J(\chi)\left(\mathcal{K}_J^{(0)}+\mathcal{K}_J^{(2)}\partial_\chi^2\right)\varphi_J(\chi)\,.
\ee
An important feature of GFT models is that the ($J$-dependent) couplings $\mathcal{K}_J^{(0)}$ and $\mathcal{K}_J^{(2)}$ can take either sign\footnote{In all interesting applications, couplings only depend on representation labels $j_I$, not on magnetic indices or intertwiners.}; in particular different modes can have couplings of different signs, so that solutions to the equations of motion are either oscillatory plane waves or real (growing and decaying) exponentials. Only modes with the latter behavior lead to a realistic cosmology \cite{lowspin}, given that for solutions to the classical Friedmann equations the volume grows or decays exponentially with respect to the scalar field. Here this property poses an immediate challenge to defining the action in Fourier space: depending on the relative signs of $\mathcal{K}_J^{(0)}$ and $\mathcal{K}_J^{(2)}$, if one wants to work with a function space that contains at least the classical solutions the notion of Fourier transform requires some careful thought.

Let us define $\mathfrak{J}^{\mathbb{C}}$ to be the space of multi-indices $J=(j_I,m_I,\iota)$ such that $\mathcal{K}_J^{(0)}$ and $\mathcal{K}_J^{(2)}$ have the same sign, and $\mathfrak{J}^{\mathbb{R}}$ to be the space of multi-indices such that $\mathcal{K}_J^{(0)}$ and $\mathcal{K}_J^{(2)}$ have opposite signs.\footnote{This notation is supposed to remind the reader of whether the classical solutions are complex or real exponentials.} We exclude the cases in which $\mathcal{K}_J^{(0)}$ or $\mathcal{K}_J^{(2)}$ vanish, which require different treatment.

For $J_1\in\mathfrak{J}^{\mathbb{C}}$ we then define
\be
\varphi_{J_1}(\chi) = \int \frac{\dd p}{2\pi}e^{\im p\chi}\varphi_{J_1}(p)\,,\quad  \bar\varphi_{J_1}(\chi) = \int \frac{\dd p}{2\pi}e^{-\im p\chi}\bar\varphi_{J_1}(p)
\ee
and the free action for such modes becomes
\be
S_{{\rm f}}^{J_1}[\varphi,\bar\varphi] = \int \frac{\dd p}{2\pi}\;\bar\varphi_J(p)\left(\mathcal{K}_J^{(0)}-\mathcal{K}_J^{(2)}p^2\right)\varphi_J(p)
\ee
which is of the form of a Klein--Gordon action in $(0+1)$ dimensions. The equations of motion are
\be
\left(\mathcal{K}_J^{(0)}-\mathcal{K}_J^{(2)}p^2\right)\varphi_J(p)=\left(\mathcal{K}_J^{(0)}-\mathcal{K}_J^{(2)}p^2\right)\bar\varphi_J(p)=0\,.
\label{eom1}
\ee

For $J_2\in\mathfrak{J}^{\mathbb{R}}$ we would like to define
\be
\varphi_{J_2}(\chi) = \int \frac{\dd P}{2\pi}e^{P\chi}\varphi_{J_2}(P)
\label{laplacetransf}
\ee
so that the field is composed of {\em real} exponential modes. But such a formula is difficult to invert; it can at best be seen as defining a two-sided Laplace transform whose inversion requires continuation of $\chi$ into the complex plane. Indeed, for imaginary $\chi=\im X$ Eq.~(\ref{laplacetransf}) would be the standard Fourier transform in $X$, which can be inverted.

Defining the free GFT action for modes in $\mathfrak{J}^{\mathbb{R}}$ requires analytic continuation in the matter field parameter $\chi$. In quantum field theory analytic continuation relies on analyticity of the quantum fields, which we need to assume here as well. For a complex scalar field {\em both} the field $\varphi$ and its conjugate $\bar\varphi$ need to be analytic in $\chi$. This implies that they cannot actually be complex conjugates for all values of $\chi$ (since then one would necessarily need to be anti-holomorphic). Our convention will be that {\em $\varphi$ and $\bar\varphi$ are complex conjugates for real values of $\chi$}. The conjugate field must then be defined as
\be
\bar\varphi_{J_2}(\chi) = \int \frac{\dd P}{2\pi}e^{P\chi}\bar\varphi_{J_2}(P)
\label{laplacetransf2}
\ee
if we also assume that the fields in $P$ space are complex conjugates of each other. 

With $\chi=\im X$, our analytic continuation prescription for the free GFT action for such modes is then
\bes
S_{{\rm f}}^{J_2}[\varphi,\bar\varphi]&=&\im\int \dd X\;\bar\varphi_J^E(X)\left(\mathcal{K}_J^{(0)}-\mathcal{K}_J^{(2)}\partial_X^2\right)\varphi_J^E(X)\nonumber
\\&=&\im\int \frac{\dd P}{2\pi}\;\bar\varphi_J(-P)\left(\mathcal{K}_J^{(0)}+\mathcal{K}_J^{(2)}P^2\right)\varphi_J(P)\,.\qquad
\ees
As usual in analytic continuation, this action is defined in terms of a ``Euclidean field'' $\varphi_J^E$ defined by $\varphi_J^E(X)=\varphi_J(\im X)$, and the action becomes purely imaginary. We have used the definition (\ref{laplacetransf}) and (\ref{laplacetransf2}) of $\varphi_J(P)$, which one may then regard as the primary definition of the GFT field modes. The action leads to the equations of motion
\be
\left(\mathcal{K}_J^{(0)}+\mathcal{K}_J^{(2)}P^2\right)\varphi_J(P) = \left(\mathcal{K}_J^{(0)}+\mathcal{K}_J^{(2)}P^2\right)\bar\varphi_J(P)=0\,.
\label{eom2}
\ee
These can now be solved mode for mode as in the case of standard relativistic field equations. When defined in Fourier space all modes now have a very similar type of dynamics. We must keep in mind that, when transforming back to $\chi$ to define time-dependent observables, an analytic continuation is needed to go from the Euclidean field defined in terms of $X$ to the ``Lorentzian field'' defined in terms of the original $\chi$. This is similar to constructions in axiomatic quantum field theory \cite{ostschr}; the field in momentum space is the primary object, which can be transformed into either real or imaginary time by applying two different types of Fourier transformations, related by analytic continuation.

The total free GFT action can then be written as\footnote{The imaginary part does not have any obvious boundedness properties so that the sign of the Wick rotation and hence the sign of the imaginary part are somewhat arbitrary.}
\bes
S_{{\rm f}}[\varphi,\bar\varphi]&=&\sum_{J\in \mathfrak{J}^{\mathbb{C}}}\int \frac{\dd p}{2\pi}\;\bar\varphi_J(p)\left(\mathcal{K}_J^{(0)}-\mathcal{K}_J^{(2)}p^2\right)\varphi_J(p)
\\&& + \sum_{J\in \mathfrak{J}^{\mathbb{R}}}\im\int \frac{\dd P}{2\pi}\;\bar\varphi_J(-P)\left(\mathcal{K}_J^{(0)}+\mathcal{K}_J^{(2)}P^2\right)\varphi_J(P)\nonumber
\ees
and has, at least in the general case, a real and an imaginary part. We now want to define a new GFT action similar to what we did in Eq.~(\ref{newaction}) for Klein--Gordon theory, adding an additional ``proper time'' parameter $\tau$ to the arguments of the GFT field. Our proposal is
\bes
S_{\ast}[\varphi,\bar\varphi,N]&=&\sum_{J\in \mathfrak{J}^{\mathbb{C}}}\int \frac{\dd p}{2\pi}\,\dd \tau\Big[\frac{\im}{2}\left(\overline\varphi_J\frac{\partial\varphi_J}{\partial \tau}-\varphi_J\frac{\partial\overline\varphi_J}{\partial \tau}\right)\nonumber
\\&&+N\left(\mathcal{K}_J^{(0)}-\mathcal{K}_J^{(2)}p^2\right)|\varphi_J|^2\Big]\nonumber
\\&&+\sum_{J\in \mathfrak{J}^{\mathbb{R}}}\int \frac{\dd P}{2\pi}\,\dd \tau\Big[\frac{\im}{2}\left(\overline\varphi_J\frac{\partial\varphi_J}{\partial \tau}-\varphi_J\frac{\partial\overline\varphi_J}{\partial \tau}\right)\nonumber
\\&&+N\left(\mathcal{K}_J^{(0)}+\mathcal{K}_J^{(2)}P^2\right)|\varphi_J|^2\Big]
\label{frozengft}
\ees
where we again stress that the fields, for each mode $J$, are now functions of both $p$ or $P$ and $\tau$. As for the Klein--Gordon field, the equations of motion then require the $\tau$ dependence to be trivial since the fields also need to satisfy Eqs.~(\ref{eom1}) and (\ref{eom2}). In this sense, the classical theory is equivalent to the one defined by Eq.~(\ref{freeakshn}). However, again as before, the ``frozen GFT'' action (\ref{frozengft}) admits a more straightforward (Dirac) canonical quantization:  the field operators in the canonical formalism should satisfy the timeless commutation relations Eq.~(\ref{timelessccr}) or
\bes
[\hat\varphi_J(p),\hat\varphi^\dagger_{J'}(p')]&=& 2\pi\delta_{J,J'}\delta(p-p')\quad(J\in\mathfrak{J}^{\mathbb{C}})\,,
\cr [ \hat\varphi_J(P),\hat\varphi^\dagger_{J'}(P') ] &=& 2\pi\delta_{J,J'}\delta(P-P')\quad(J\in\mathfrak{J}^{\mathbb{R}})\,,
\label{newtimelessccr}
\ees
and thus again generate a kinematical Fock space, equivalent to the one used in GFT in the timeless setting. The constraint on a state $|\psi\rangle$ to be physical is then
\bes
\left(\mathcal{K}_J^{(0)}-\mathcal{K}_J^{(2)}p^2\right)\hat\varphi^\dagger_{J}(p)\hat\varphi_J(p)|\psi\rangle &=&\nonumber
\\ \left(\mathcal{K}_J^{(0)}+\mathcal{K}_J^{(2)}P^2\right)\hat\varphi^\dagger_{J}(P)\hat\varphi_J(P)|\psi\rangle &=& 0\,.
\ees
The fact that we impose these constraints strongly is our departure from previous work in the timeless formalism.
Again, $\hat\varphi^\dagger_{J}(p)\hat\varphi_J(p)$ and $\hat\varphi^\dagger_{J}(P)\hat\varphi_J(P)$ are number densities on the Fock space and these constraints imply that only modes satisfying the constraints
\be
\left(\mathcal{K}_J^{(0)}-\mathcal{K}_J^{(2)}p^2\right)=0\,,\quad \left(\mathcal{K}_J^{(0)}+\mathcal{K}_J^{(2)}P^2\right)=0
\ee
can be excited for a state to be considered physical. With the shorthands
\be
\mu_J=\sqrt{\frac{\mathcal{K}_J^{(0)}}{\mathcal{K}_J^{(2)}}}\quad(J\in\mathfrak{J}^{\mathbb{C}})\,,\quad m_J=\sqrt{\frac{-\mathcal{K}_J^{(0)}}{\mathcal{K}_J^{(2)}}}\quad(J\in\mathfrak{J}^{\mathbb{R}})
\ee
these constraints become more simply $\mu_J^2-p^2=0$ and $m_J^2-P^2=0$. It should then be clear that the entire discussion of Sec.~\ref{frozensec} can be extended to the case of GFT: there exists a map
\be
\hat\varphi_J(p)\mapsto \mathcal{P}^{\mathbb{C}}\hat\varphi_J(p)\,,\quad \hat\varphi^\dagger(p)\mapsto \mathcal{P}^{\mathbb{C}}\hat\varphi^\dagger(p)
\ee
and a similar map $\mathcal{P}^{\mathbb{R}}$ for the $P$ modes, such that the projected operators generate a physical Fock space whose one-particle sector is the physical Hilbert space one would construct through group averaging. If we define
\be
\mathcal{P}^{\mathbb{C}}\hat\varphi_J^\dagger(p)=\frac{\sqrt{2\pi}}{\sqrt[4]{2\mathcal{K}_J^{(0)}\mathcal{K}_J^{(2)}}}\left(\delta(p-\mu_J)\hat{a}_J^\dagger+\delta(p+\mu_J)\hat{b}_J^\dagger\right)
\label{projp}
\ee
for $J\in\mathfrak{J}^{\mathbb{C}}$ and
\be
\mathcal{P}^{\mathbb{R}}\hat\varphi_J^\dagger(P)=\frac{\sqrt{2\pi}}{\sqrt[4]{-2\mathcal{K}_J^{(0)}\mathcal{K}_J^{(2)}}}\left(\delta(P-m_J)\hat{A}_J^\dagger+\delta(P+m_J)\hat{B}_J^\dagger\right)
\label{projP}
\ee
for $J\in\mathfrak{J}^{\mathbb{R}}$ where $\hat{a}_J,\hat{a}_J^\dagger$ and the three other canonical pairs satisfy the usual algebra of creation and annihilation operators, {\em i.e.},
\be
[\hat{a}_J,\hat{a}^\dagger_{J'}]=\delta_{J,J'}\,,
\ee
then the inner product between physical single-particle Fock states associated to a $J\in \mathfrak{J}^{\mathbb{C}}$ mode defined by
\be
|\psi_{{\rm ph}}\rangle = \int \frac{\dd p}{2\pi}\psi(p)\mathcal{P}^{\mathbb{C}}\hat\varphi_J^\dagger(p)|0\rangle_{{\rm ph}}
\ee
is
\bes
\langle \phi_{{\rm ph}}|\psi_{{\rm ph}} \rangle &=& \frac{1}{2\pi\sqrt{2\mathcal{K}_J^{(0)}\mathcal{K}_J^{(2)}}}\left(\overline{\phi(\mu_J)}\psi(\mu_J)\right.\nonumber
\\&&+\left.\overline{\phi(-\mu_J)}\psi(-\mu_J)\right)\nonumber
\\&=&\int \frac{\dd p}{2\pi}\,\delta\left(\mathcal{K}_J^{(0)}-\mathcal{K}_J^{(2)}p^2\right)\,\overline{\phi(p)}\psi(p)\qquad
\ees
in agreement with group averaging. The same is true for the modes $J\in\mathfrak{J}^{\mathbb{R}}$ corresponding to real exponential solutions; the calculation is the same up to a minus sign.

Things become more interesting if we consider the dependence of physical states and observables on $\chi$. We can again insert a time evolution factor $e^{\im p \chi}$ into the map (\ref{projp}) to obtain physical states defined in the Schr\"odinger picture at arbitrary $\chi$. This yields
\bes
\mathcal{P}^{\mathbb{C}}_\chi\hat\varphi_J^\dagger(p)&=&\frac{\sqrt{2\pi}}{\sqrt[4]{2\mathcal{K}_J^{(0)}\mathcal{K}_J^{(2)}}}\left(e^{\im\mu_J\chi}\delta(p-\mu_J)\hat{a}_J^\dagger+\right.\nonumber
\\&&\left.e^{-\im\mu_J\chi}\delta(p+\mu_J)\hat{b}_J^\dagger\right)
\ees
in analogy to Eq.~(\ref{timedepproj}) in the case of Klein--Gordon theory. For the $P$ modes we must remember the need for analytic continuation: the evolution operator is of the form $e^{\im P X}=e^{P \chi}$, a real exponential when expressed in terms of $\chi$. Hence the $\chi$-dependent version of Eq.~(\ref{projP}) is
\bes
\mathcal{P}^{\mathbb{R}}_\chi\hat\varphi_J^\dagger(P)&=&\frac{\sqrt{2\pi}}{\sqrt[4]{-2\mathcal{K}_J^{(0)}\mathcal{K}_J^{(2)}}}\left(e^{m_J\chi}\delta(P-m_J)\hat{A}_J^\dagger+\right.\nonumber
\\&&\left.e^{-m_J\chi}\delta(P+m_J)\hat{B}_J^\dagger\right)
\ees
in accordance with the classical solutions to the GFT field equations for these modes, which are real (growing and decaying) exponentials. Since we require $\hat\varphi$ and $\hat\varphi^\dagger$ to be Hermitian conjugates for real $\chi$ arguments the corresponding projection for $\hat\varphi$ is
\bes
\mathcal{P}^{\mathbb{R}}_\chi\hat\varphi_J(P)&=&\frac{\sqrt{2\pi}}{\sqrt[4]{-2\mathcal{K}_J^{(0)}\mathcal{K}_J^{(2)}}}\left(e^{m_J\chi}\delta(P-m_J)\hat{A}_J+\right.\nonumber
\\&&\left.e^{-m_J\chi}\delta(P+m_J)\hat{B}_J\right)
\ees
so that, unlike for the $p$ modes, the $\chi$-dependent exponential factors do not switch sign between $\hat\varphi_J$ and $\hat\varphi^\dagger_J$. This is as it should be: for the oscillatory $p$ modes, for each of the two (positive or negative frequency) solutions  $\hat\varphi_J^\dagger\hat\varphi_J$ should be time-independent for each mode, given that the classical solutions are plane waves for which $|\varphi_J|^2$ is a constant. For the $P$ modes, this is not the case and so the combination $ \hat\varphi_J^\dagger\hat\varphi_J$ should {\em not} be time-independent.

We can now look at GFT observables. On the kinematical Hilbert space, one class of Dirac observables is of the form
\bes
\hat{O}_f &=& \sum_{J\in\mathfrak{J}^{\mathbb{C}}} \int \frac{\dd p}{2\pi}f_J(p)\hat\varphi_J^\dagger(p)\hat\varphi_J(p)+\nonumber
\\&&\sum_{J\in\mathfrak{J}^{\mathbb{R}}} \int \frac{\dd P}{2\pi}F_J(P)\hat\varphi_J^\dagger(P)\hat\varphi_J(P)\,;
\label{gftobserv1}
\ees
these observables preserve the space of physical states since they do not excite any unphysical modes. If we again define these as time-dependent operators in the Heisenberg picture, the required projection for bilinear operators is
\bes
&&\mathcal{P}^{\mathbb{C}}_\chi\left(\hat\varphi_J^\dagger(p)\hat\varphi_J(p)\right)
\label{gftproj1}
\\&=& 2\pi\left(\delta(p-\mu_J)\hat{a}_J^\dagger\hat{a}_J+\delta(p+\mu_J)\hat{b}_J^\dagger\hat{b}_J\right)\nonumber
\ees
for $p$ modes, where the normalization is again fixed by requiring these to count particles in integer amounts. However, for $P$ modes corresponding to $J\in\mathfrak{J}^{\mathbb{R}}$ we have
\bes
&&\mathcal{P}^{\mathbb{R}}_\chi\left(\hat\varphi_J^\dagger(P)\hat\varphi_J(P)\right)
\label{gftproj2}
\\&=& 2\pi\left(e^{2m_J\chi}\delta(P-m_J)\hat{A}_J^\dagger\hat{A}_J+e^{-2m_J\chi}\delta(P+m_J)\hat{B}_J^\dagger\hat{B}_J\right)\nonumber
\ees
again by demanding that expectation values in the Heisenberg and Schr\"odinger picture agree. Therefore, on the physical Hilbert space the total number of particles in the $P$ modes takes the form
\be
\hat{N}_{{\rm ph}}^{P} = \sum_{J\in\mathfrak{J}^{\mathbb{R}}} \left(e^{2m_J\chi}\hat{A}_J^\dagger\hat{A}_J+e^{-2m_J\chi}\hat{B}_J^\dagger\hat{B}_J\right)
\ee
in contrast with the total number of $p$ particles
\be
\hat{N}_{{\rm ph}}^{p} = \sum_{J\in\mathfrak{J}^{\mathbb{C}}} \left(\hat{a}_J^\dagger\hat{a}_J+\hat{b}_J^\dagger\hat{b}_J\right)
\ee
which is independent of time, as it was in the previous case of conventional Klein--Gordon theory, cf.~Eq.~(\ref{physnumb}).

Conversely, observables can be time-dependent for $p$ modes but time-independent for $P$ modes. Indeed, in analogy with Eq.~(\ref{weirdob}) consider
\bes
\hat{Q}_f &=&  \sum_{J\in\mathfrak{J}^{\mathbb{C}}} \int \frac{\dd p}{2\pi}f_J(p)\hat\varphi_J^\dagger(p)\hat\varphi_J(-p) + \nonumber
\\&& \sum_{J\in\mathfrak{J}^{\mathbb{R}}} \int \frac{\dd P}{2\pi}F_J(P)\hat\varphi_J^\dagger(P)\hat\varphi_J(-P)
\label{weirdob2}
\ees
(with $f_J(p)=\overline{f_J(-p)}$ and $F_J(p)=\overline{F_J(-p)}$) as observables on the kinematical Hilbert space; again these do not excite any unphysical modes and thus preserve the physical Hilbert space. The relevant (time-dependent) map acting on the bilinears appearing in Eq.~(\ref{weirdob2}) is
\bes
&&\mathcal{P}^{\mathbb{C}}_\chi\left(\hat\varphi_J^\dagger(p)\hat\varphi_J(-p)\right)
\\&=& 2\pi\left(e^{-2\im\mu_J\chi}\delta(p-\mu_J)\hat{a}_J^\dagger\hat{b}_J+e^{2\im\mu_J\chi}\delta(p+\mu_J)\hat{b}_J^\dagger\hat{a}_J\right)\nonumber
\ees
but
\bes
&&\mathcal{P}^{\mathbb{R}}_\chi\left(\hat\varphi_J^\dagger(P)\hat\varphi_J(-P)\right)
\\&=& 2\pi\left(\delta(P-m_J)\hat{A}_J^\dagger\hat{B}_J+\delta(P+m_J)\hat{B}_J^\dagger\hat{A}_J\right)\nonumber
\ees
so that, for $f_J(p)=0$, one obtains a time-independent observable on the physical Hilbert space:
\be
\mathcal{P}_\chi \hat{Q}_f = \sum_{J\in\mathfrak{J}^{\mathbb{R}}} \left(F_J(m_J)\hat{A}_J^\dagger\hat{B}_J+\overline{F_J(m_J)}\hat{B}_J^\dagger\hat{A}_J\right)\,.
\ee
These constructions define a physical Hilbert space for GFT, obtained from a frozen formalism and Dirac-type quantization as previously defined for the Klein--Gordon field, together with a set of physical observables.

The physical Hilbert space is a Fock space in which each Peter--Weyl mode $J$ is associated with two creation operators, either $\hat{a}^\dagger_J$ and $\hat{b}^\dagger_J$ or $\hat{A}^\dagger_J$ and $\hat{B}^\dagger_J$, and thus two types of excitations which one may consider as analogous to particle and antiparticle. This Fock space is the direct sum of two copies of the Fock space constructed in the deparametrized setting of Ref.~\cite{edham} and further studied, {\em e.g.}, in Ref.~\cite{generalcosmology}. This doubling of degrees of freedom is due to the fact that we are considering a complex GFT field whereas the previous works in the deparametrized setting focused on the case of a {\em real} field. It is a curious feature of the frozen formalism we have introduced that it only straightforwardly applies to complex fields, due to the need for two basic operators to be defined as canonically conjugate on the kinematical Hilbert space.

The physical Hilbert space of this GFT quantization is much smaller than the  kinematical Hilbert space generated by the initial field operators defined by Eq.~(\ref{newtimelessccr}). We have identified maps from the kinematical to the physical Hilbert space, which remove all $p$ and $P$ modes apart from the ones satisfying Eqs.~(\ref{eom1}) or (\ref{eom2}). These constraints were imposed strongly, not weakly as previously in the timeless quantization of GFT.

\section{Relational observables and effective cosmology}
\label{relateff}

The main application of the operator formalism for GFT has been the derivation of effective cosmological dynamics from the fundamental theory \cite{QCGFT,GFTcond}. This derivation makes crucial use of {\em relational observables} whose expectation values are computed for a particular class of states, leading to effective dynamics written in terms of these observables. Effective cosmological dynamics have been derived in the timeless setting in a mean-field approximation, but also in the deparametrized approach \cite{edham,generalcosmology}. The effective cosmology obtained in both settings has similar properties: the dynamics reduce to the classical Friedmann equations at large volume but there are high-curvature corrections which lead to a bounce interpolating between the classical collapsing and expanding solutions. The details of these corrections are slightly different between the different approaches. 

The most important relational observable in the timeless GFT setting defines the total volume (of space) at a given value of ``relational time'' $\chi$. This observable was introduced in Ref.~\cite{QCGFT} and mimics the analogous observable used to characterize the dynamics of the Universe in loop quantum cosmology \cite{LQC}. 

In the notation used in this paper, this relational volume observable on the kinematical Hilbert space is 
\be
\hat{V}(\chi) = \sum_J v_J \,\hat\varphi_J^\dagger(\chi)\hat\varphi_J(\chi)
\label{vchi}
\ee
where $v_J$ is the volume eigenvalue (``volume per GFT quantum'') associated to the representation $J$. The meaning of such an observable on the kinematical GFT Hilbert space is somewhat murky, given that this Hilbert space does not contain a subspace of modes at fixed $\chi$; normalizable states must be, {\em e.g.}, wavepackets containing different values of $\chi$. As a result $\hat{V}(\chi)$ should really be considered as a density to be ``smeared'' over a finite $\chi$ range, as already discussed in Ref.~\cite{QCGFT} and in more detail in Refs.~\cite{GFTpert,mehdiisha2,lucadaniele}. This does not necessarily affect the resulting cosmology expressed in terms of expectation values of $\hat{V}(\chi)$, but causes issues when higher moments, {\em i.e.}, quantum fluctuations are considered.

A second observable used in Ref.~\cite{QCGFT} is given by
\be
\hat{\pi}_\chi = \sum_J -\frac{\im}{2}\left(\hat\varphi_J^\dagger(\chi)\partial_\chi\hat\varphi_J(\chi)-\Big(\partial_\chi\hat\varphi_J^\dagger(\chi)\Big)\hat\varphi_J(\chi)\right)
\label{pichi}
\ee
and identified with the conjugate momentum to the scalar field $\chi$, in analogy with usual arguments in quantum mechanics (thinking of the canonical momentum as a generator of translations in $\chi$). $\pi_\chi$ is conserved in the classical and timeless quantum theory; however its conservation is due to a global ${\rm U}(1)$ symmetry of the theory and at least {\em a priori} unrelated to translations in $\chi$ \cite{QCGFT} (see also Ref.~\cite{multiplefields} for why $\pi_\chi$ is not the conjugate momentum to $\chi$). The conservation law for $\hat{\pi}_\chi$ is important in the cosmological interpretation of the theory since it justifies identifying this quantity with a conserved momentum in cosmology, but it seems one should define a smeared momentum and perhaps a smeared conservation law, which has not been done.

The conceptual issues with Eqs.~(\ref{vchi})--(\ref{pichi}) are connected to the fact that one has defined these observables on the kinematical Hilbert space generated by Eq.~(\ref{timelessccr}); they are defined in relational terms but evaluated in the kinematical inner product. Using the projection maps acting on operators on the kinematical Hilbert space we can now define the equivalent of such observables on a {\em physical} Hilbert space. This is where observables are defined in a Dirac-quantized theory such as, {\em e.g.}, in loop quantum cosmology \cite{LQC}.

The first observable defining a relational volume is obtained straightforwardly from the general expression (\ref{gftobserv1}) by setting $f_J(p)=F_J(p)=v_J$, {\em i.e.}, by choosing the corresponding kinematical observable to be the total volume of {\em all} particles. Applying the maps (\ref{gftproj1}) and (\ref{gftproj2}) to this operator we find the corresponding observable on the physical Hilbert space
\bes
\hat{V}_{{\rm ph}} &=& \sum_{J\in\mathfrak{J}^{\mathbb{C}}} v_J\left(\hat{a}_J^\dagger\hat{a}_J+\hat{b}_J^\dagger\hat{b}_J\right)
\label{newvolume}
\\&& +\sum_{J\in\mathfrak{J}^{\mathbb{R}}} v_J\left(e^{2m_J\chi}\hat{A}_J^\dagger\hat{A}_J+e^{-2m_J\chi}\hat{B}_J^\dagger\hat{B}_J\right)\nonumber
\ees
which is essentially of the form of the relational volume observable defined in Eq.~(\ref{vchi}). Rather than defining ``by hand'' relational observables at a fixed value of $\chi$, in the frozen formalism {\em all} observables are naturally of this form since one can think of the physical Hilbert space as defined at a given value of $\chi$. This is perhaps most explicit in the Heisenberg picture in which one can think of (Dirac) observables as evolving in $\chi$. Eq.~(\ref{newvolume}) shows that the total volume of the oscillatory modes is constant in $\chi$ whereas the volume of the real exponential modes has an exponentially growing and an exponentially decaying piece, in line with the behavior of classical solutions.

To obtain an observable analogous to the ${\rm U}(1)$ charge (\ref{pichi}), we integrate the classical version of Eq.~(\ref{pichi}) over $\chi$ and apply the Fourier transform defined in Sec.~\ref{frozengftsec}. The corresponding operator on the kinematical Hilbert space is then
\bes
\hat\Pi &=& \sum_{J\in\mathfrak{J}^{\mathbb{C}}} \int \frac{\dd p}{2\pi}\,p\,\hat\varphi^\dagger_J(p)\hat\varphi_J(p)\nonumber
\\&& -\im \sum_{J\in\mathfrak{J}^{\mathbb{R}}} \int \frac{\dd P}{2\pi}\,P\,\hat\varphi^\dagger_J(P)\hat\varphi_J(-P)\,;
\label{piobs}
\ees
this is mapped to an observable on the physical Hilbert space that is conserved for each mode separately,
\bes
\hat\Pi_{{\rm ph}} &=& \sum_{J\in\mathfrak{J}^{\mathbb{C}}} \mu_J\,\left(\hat{a}_J^\dagger\hat{a}_J-\hat{b}_J^\dagger\hat{b}_J\right)\nonumber
\\&& -\im \sum_{J\in\mathfrak{J}^{\mathbb{R}}} m_J\left(\hat{A}_J^\dagger\hat{B}_J-\hat{B}_J^\dagger\hat{A}_J\right)\,.
\label{newpi}
\ees
Eq.~(\ref{newpi}) corresponds to  a quantization of classically conserved quantities: writing a classical oscillatory solution as $\varphi=\alpha e^{\im\mu\chi}+\beta e^{-\im\mu\chi}$ and a real exponential solution as $\varphi=A e^{m\chi}+B e^{-m\chi}$, the conserved quantities associated to the ${\rm U}(1)$ symmetry of the theory are $\mu(|\alpha|^2-|\beta|^2)$ for the first and $-\im m(\bar{A}B-\bar{B}A)$ for the second (see, {\em e.g.}, Ref.~\cite{lowspin}). This is obviously also the expression one finds when evaluating Eq.~(\ref{pichi}) in a mean-field approximation. $\hat\Pi_{{\rm ph}}$ is relational in the sense that one can think of it as a $\chi$-dependent observable which happens to be a constant of motion. Notice that the expectation value of Eq.~(\ref{piobs}) would either be zero or divergent on any physical state, just as classically the integral of a conserved quantity over time is either zero or diverges.

It is clear that one can similarly define observables on the physical Hilbert space from any well-defined (Dirac) observable on the kinematical Hilbert space. The resulting physical observables are generally time-dependent (in the Heisenberg picture) but their fluctuations and higher $n$-point functions are regular functions in $\chi$ and do not encounter the divergences seen in Refs.~\cite{GFTpert,mehdiisha2}. They can then be used to define an effective cosmology in analogy with previous work in GFT \cite{QCGFT,edham,generalcosmology}. 

We can illustrate this by focusing on the simplest case in which one considers only a single $J$ mode with real exponential solutions; then from Eq.~(\ref{newvolume}) we have
\be
\left(\frac{1}{\langle \hat{V}_{{\rm ph}}\rangle}\frac{\dd \langle\hat{V}_{{\rm ph}}\rangle}{\dd \chi}\right)^2 = 4m_J^2 \left(1-4\frac{v_J^2\langle\hat{A}_J^\dagger\hat{A}_J\rangle\langle\hat{B}_J^\dagger\hat{B}_J\rangle}{\langle \hat{V}_{{\rm ph}}\rangle^2}\right)
\label{effectivefriedmann}
\ee
as our effective Friedmann equation. The effective cosmology has the general features previously found both in the timeless and deparametrized settings \cite{QCGFT,edham,generalcosmology}: if the GFT couplings are such that $m_J^2=3\pi G$ with $G$ the low-energy Newton's constant, at large volumes the dynamics reduce to the classical Friedmann equation
\be
\left(\frac{1}{V_{{\rm GR}}}\frac{\dd V_{{\rm GR}}}{\dd \chi}\right)^2 = 12\pi G\,.
\ee
The correction to the the classical Friedmann equation appearing in Eq.~(\ref{effectivefriedmann}) scales as $1/V^2$ and hence can be written as $-\rho/\rho_c$ if one identifies $\rho=M/V^2$, where $M$ is a positive constant, with the classical energy density of a massless scalar field and where $\rho_c$ is a constant. Such a term leads to a bounce when the energy density reaches $\rho=\rho_c$, just as it does in loop quantum cosmology. (To make this identification more precise we would need to identify a specific relation between the occupation numbers $\langle\hat{A}_J^\dagger\hat{A}_J\rangle$ and $\langle\hat{B}_J^\dagger\hat{B}_J\rangle$ and the constant $M$.) Only very special initial conditions such that $\langle\hat{A}_J^\dagger\hat{A}_J\rangle=0$ or $\langle\hat{B}_J^\dagger\hat{B}_J\rangle=0$ do not feature such a bounce; indeed, from Eq.~(\ref{newvolume}) it is clear that such states follow exactly either the classical contracting or the expanding solution. We do not see a second correction term scaling as $1/V$, as found in Ref.~\cite{QCGFT}, which can be traced back to the fact that dynamics were not imposed strongly in Ref.~\cite{QCGFT}.

Operators on the kinematical Hilbert space that are not well-defined on physical states cannot be given a clear definition within the frozen formalism. The most important example of this is a would-be operator corresponding to the massless scalar field $\chi$, which after Fourier transform becomes
\bes
\hat\chi &=& \im\sum_{J\in\mathfrak{J}^{\mathbb{C}}} \int \frac{\dd p}{2\pi}\,\hat\varphi^\dagger_J(p)\frac{\dd\hat\varphi_J(p)}{\dd p}\nonumber
\\&& + \sum_{J\in\mathfrak{J}^{\mathbb{R}}} \int \frac{\dd P}{2\pi}\,\hat\varphi^\dagger_J(-P)\frac{\dd\hat\varphi_J(P)}{\dd P}\,.
\label{chiop}
\ees
There seems to be no straightforward way to make sense of the derivative of a field operator on the physical Hilbert space, given that for each $J$ only two isolated values of $p$ or $P$ correspond to physical states. This observation is not surprising: it is the GFT equivalent of the statement that there is no time observable in quantum mechanics or quantum field theory, which goes back to Pauli \cite{pauli}. Pauli's statement relies on having a Hamiltonian that is bounded from below, which may not exist in GFT; Hamiltonians constructed in the deparametrized setting are unbounded \cite{edham}. The observation does not imply that there is no useful notion of ``time'' observable in the frozen formalism for GFT. Indeed, given that Eq.~(\ref{chiop}) does not work, one could try to  construct more general notions of time observable, {\em e.g.}, by using the sophisticated machinery of positive operator valued measures \cite{trinitystuff}. It would be interesting to find an explicit construction of this type in GFT.

\section{Conclusions}

We have proposed a new perspective on the canonical quantization of quantum field theories: we have suggested an action for a complex scalar field in which fields depend on a parameter $\tau$, which can be seen as the analog of proper time for a relativistic particle. Dynamical equations force the fields to be independent of $\tau$. We call this a frozen formalism in analogy with the quantum dynamics of a relativistic particle for which reparametrization invariance forces states and observables to be independent of $\tau$. Dirac quantization of this new action leads to a kinematical Fock space and, after the imposition of constraints, a physical Hilbert space equivalent to the usual Fock space. We have introduced ``projection'' maps from states and (Dirac) observables on the kinematical to those on the physical Hilbert space.

We then applied this frozen formalism to group field theory models for quantum gravity coupled to a massless scalar field $\chi$, showing how the kinematical Hilbert space of the frozen formalism is equivalent to the Hilbert space proposed in a ``timeless'' quantization in the literature. Imposing the constraints strongly and not weakly as was done previously, we  obtain a physical Hilbert space which is equivalent to one found through a different, deparametrized canonical quantization. Thus the frozen formalism links between the timeless and deparametrized approaches and shows in particular how physical observables can be defined on the physical Hilbert space, taking into account the fact well-known from Dirac quantization that the inner products on kinematical and physical Hilbert spaces cannot be assumed to be the same. This more careful construction avoids some of the pathologies encountered for the timeless formalism in previous work. It does not alter the main results for effective cosmology, which are based on expectation values only.

In this paper, we restricted ourselves to quadratic GFT actions, which made it straightforward to implement a Dirac quantization and construct the physical Fock space. Models of physical interest for quantum gravity include higher order terms which make an explicit construction of the physical Hilbert space for such theories very difficult. In this more general case, it may then be necessary to accept that the  dynamics are solved only approximately, such as in expectation values, as often suggested for GFT in the timeless formalism \cite{LQGFT,GFTcond}. Also the inner product and physical observables may then only be known approximately. This situation for the canonical quantization of GFT can be compared to canonical LQG, where implementing group averaging to construct a physical inner product from the full Hamiltonian constraint of general relativity is not straightforwardly possible and serious efforts have focused on developing novel techniques, {\em e.g.}, Ref.~\cite{LQGmethods}. While the relatively simple scalar field theory structure of GFT may make the task look easier than for LQG, only future work can show whether there are any prospects for achieving an exact Dirac quantization for an interacting GFT. Nevertheless, it seems that the general lessons drawn from the quadratic case have to hold also in this situation: physical and kinematical Hilbert space cannot be identified and not all Hermitian operators on the kinematical Hilbert space are physical observables. The meaning of calculations done on the kinematical Hilbert space can be obscure, as in LQG \cite{discretespec}. It would be important to understand how the constructions in this paper can be extended to a more general setting in an approximate sense. Effective methods such as developed in Ref.~\cite{effprob}, which do not require knowledge of the full Hilbert space but use expectation values and higher moments, may also be useful. In contrast, the deparametrized approach can be extended more straightforwardly to interacting models, given that the scalar $\chi$ remains a good clock also in this more general case. A simple interaction term has been studied in Ref.~\cite{generalcosmology}. However, it is unclear whether the deparametrized approach can be generally covariant, as is already apparent when multiple scalars are coupled \cite{multiplefields}.

Some peculiarities of the frozen formalism are related to general properties of the Dirac quantization of relativistic systems. The formalism we defined can only be straightforwardly applied to complex scalar fields; the physical Fock space then always contains both particle and antiparticle excitations. The underlying reason for this can be understood from the Dirac quantization of a single relativistic particle, in which the physical Hilbert space constructed through group averaging contains positive {\em and} negative frequency solutions, associating a positive norm to both. This is in contrast to the standard treatment of a real Klein--Gordon field, which only treats positive frequency modes as physical excitations. Of course, one can decide to work only with particles or antiparticles, and reduce the physical Hilbert space to that of a real field theory by hand. It is not clear how one would define a Dirac quantization directly for real fields, just as it is not clear how to construct a group averaging procedure that uses only one of the two classical solutions to the constraint for the relativistic particle. In GFT, both the general perspective of Section \ref{GFTintro} and the deparametrized quantization of Ref.~\cite{edham} allow for real fields; the interpretation of GFT Feynman amplitudes as spin foam amplitudes only requires real fields \cite{SFGFT}. When complex fields are used, one ends up with two copies of the Hilbert space that would seem required from the perspective of LQG. As we mentioned below Eq.~(\ref{timedepproj}), an interaction between particle and antiparticle sectors might lead to unphysical results, so one would have to study whether a GFT that couples these two sectors has a good interpretation from the perspective of LQG. One might decide that such interactions are forbidden.

The frozen formalism proposed here is in principle  more generally applicable to quantum systems which do not have any ``time parameter'' with respect to which evolution could be defined, such as GFT models without a massless scalar field, but also more general combinatorial approaches to quantum gravity such as matrix and tensor models \cite{matrixtens}. It would be very interesting to study further applications in this direction.

{\em Acknowledgments.} --- I thank Philipp H\"ohn, Axel Polaczek and Edward Wilson-Ewing for helpful discussions and comments on the manuscript, and the referee for suggestions on extending the discussion. This work was funded by the Royal Society under a Royal Society University Research Fellowship (UF160622) and a Research Grant for Research Fellows (RGF\textbackslash R1\textbackslash 180030).

\end{document}